\documentclass[prd,twocolumn,superscriptaddress]{revtex4-1}
\usepackage{graphicx}%
\usepackage{dcolumn}
\usepackage{amsmath,amsthm,amssymb}
\begin{document}

\title{Nonlinear evolution of cylindrical gravitational waves: numerical method and physical aspects}


\author{Juliana Celestino}
\email{juliana.efei@gmail.com}
\affiliation{
Departamento de F\'{\i}sica Te\'orica, Instituto de F\'{\i}sica
A. D. Tavares, Universidade do Estado do Rio de Janeiro \\
R. S\~ao Francisco Xavier, 524. Rio de Janeiro, RJ, 20550-013, Brazil}

\author{H. P. de Oliveira}
\email{hp.deoliveira@pq.cnpq.br}
\affiliation{
Departamento de F\'{\i}sica Te\'orica, Instituto de F\'{\i}sica
A. D. Tavares, Universidade do Estado do Rio de Janeiro \\
R. S\~ao Francisco Xavier, 524. Rio de Janeiro, RJ, 20550-013, Brazil}

\affiliation{Department of Physics and Astronomy, University of Pittsburgh\\
100 Allen Hall, 3941, O'Hara St., Pittsburgh, PA 15260, USA}

\author{E. L. Rodrigues}
\email{eduardo.rodrigues@unirio.br}
\affiliation{Departamento de Ci\^encias Naturais, Instituto de Bioci\^encias, Universidade Federal do Estado do Rio de Janeiro \\
Av. Pasteur, 458. Rio de Janeiro, RJ, 22290-040, Brazil}

\date{\today}

\begin{abstract}
General cylindrical waves are the simplest axisymmetrical gravitational waves that contain both $+$ and $\times$ modes of polarization. In this paper, we have studied the evolution of general cylindrical gravitational waves in the realm of the characteristic scheme with a numerical code based on the Galerkin-Collocation method. The investigation consists of the numerical realization of concepts such as Bondi mass and the news functions adapted to cylindrical symmetry. The Bondi mass decays due to the presence of the news functions associated with both polarization modes. We have interpreted that each polarization mode as channels from which mass is extracted. Under this perspective, we have presented the enhancement effect of the polarization mode $+$ due to the nonlinear interaction with the mode $\times$. After discussing the role of matter in cylindrical symmetry, we have extended the numerical code to include electromagnetic fields.
\end{abstract}

\maketitle

\section{Introduction}%

We describe here the evolution of general cylindrical gravitational waves using a numerical code based on the Galerkin-Collocation method \cite{gar_col}. The novelty is the numerical realization of the characteristic framework established by Stachel \cite{stachel} in extending the Bondi-Sachs formalism \cite{bondi_sachs} to cylindrical symmetry. One can find concepts like the Bondi mass, the news functions, the Bondi formula governing the decay of the Bondi mass, and the asymptotic structure of the Riemann tensor that are crucial ingredients to describe properly the gravitational radiation emitted by an isolated source.

In exploring the evolution of cylindrical waves, we have exhibited the decay of the Bondi mass due to the action of the news functions. In the general case, cylindrical waves are unpolarized with two polarization modes, $+$ and $\times$, hereafter represented by the metric potentials $\psi$ and $\omega$ respectively. Each mode has the corresponding news functions that according to the Bondi formula can be understood as channels from which mass is carried out. In this direction, we have shown an interesting enhancement effect due to the nonlinear interaction between both modes. It turns to be more evident when the potential $\psi$ vanishes initially whereas the potential $\omega$ has its initial amplitude as a free parameter. For small values of this amplitude, the amount of mass extracted by the mode $\times$ is greater than the corresponding extracted by the mode $+$. However, there exists a critical value above which the mass extracted by the mode $+$ surpass the amount carried out by the mode $\times$.

The main motivation for studying cylindrical gravitational waves is their simplicity that allows to investigate the nonlinear interaction of distinct polarization modes \cite{thorne,piran,goncalves,clarke_dinverno}. Polarized cylindrical waves can be described analytically by solving the resulting linear wave equation. These waves are known as the Einstein-Rosen waves that became the first exact description of a gravitational wave \cite{ER}. Piran at al \cite{piran} presented the first numerical evolution of general cylindrical waves in the Cauchy formalism. They have exhibited the gravitational counterpart of the Faraday rotation valid for electromagnetic waves.

We have organized the paper as follows. Section II shows the basic equations of cylindrical spacetimes in the characteristic scheme including the expressions for the Bondi mass, the news functions, and the Bondi formula. We have presented the Galerkin-Collocation code in Section III, and the convergence tests are in Section IV. Section V discusses physical aspects such as the enhancement effect resulting from the nonlinear interaction of the polarization modes, and the role of matter in cylindrical symmetry. In this direction, we have extended the code to describe the dynamics of electromagnetic waves in cylindrical symmetry. The enhancement effect takes place as a consequence of the nonlinear interaction of the gravitational and electromagnetic potentials. Finally, in Section VI we summarize and trace some perspectives of the present work.

\section{Spacetimes with cylindrical symmetry}%


We have considered the general cylindrical line element originally proposed by Kompaneets \cite{kompaneets} and Jordan et al. \cite{jordan} in null coordinates,

\begin{eqnarray}
ds^2 =&& -{\rm e}^{2(\gamma - \psi)} (du^2 + 2~du~d\rho) + {\rm e}^{2\psi}(dz+\omega d\phi )^{2} + \nonumber \\
&& + \rho^{2}{\rm e}^{-2\psi}d\phi ^{2}, \label{eq1}
\end{eqnarray}

\noindent where $u$ is the retarded null coordinate that foliates the spacetime in hypersurfaces $u=\mathrm{constant}$ and $(\rho,z,\phi)$ are the usual cylindrical coordinates. The metric functions $\psi$, $\omega$, $\gamma$ depend on $u$ and $\rho$. The relevant field equations in vacuum are,

\begin{widetext}
\begin{eqnarray}
&&2\rho \psi _{,u\rho}+\psi _{,u}-\psi _{,\rho}-\rho \psi _{,\rho\rho}-\frac{{\rm e}^{4\psi}}{2\rho}(2\omega_{,u}\omega_{,\rho}-\omega_{,\rho}^{2})=0  \label{eq2}\\
\nonumber \\
&&2\omega _{,u\rho}-\omega _{,\rho\rho}+\frac{\omega_{,\rho}-\omega_{,u}}{\rho}-4(\omega_{,\rho}\psi_{,\rho}-\omega_{,\rho}\psi_{,u}-\omega_{,u}\psi_{,\rho})=0 \label{eq3} \\
\nonumber \\
&&\gamma_{,\rho}=\rho\psi_{,\rho}^{2}+\frac{{\rm e}^{4\psi}}{4\rho}\omega_{,\rho}^{2} \label{eq4} \\
\nonumber \\
&&\gamma_{,u}=2\rho(\psi _{,u}\psi _{,\rho}-\psi _{,u}^{2})+\frac{{\rm e}^{4\psi}}{2\rho}(\omega_{,u}\omega_{,\rho}-\omega_{,u}^{2}). \label{eq5}
\end{eqnarray}
\end{widetext}

\noindent The subscripts $u$ and $\rho$ denote partial derivatives with respect to these coordinates. As a well-known important aspect of cylindrical spacetimes \cite{thorne}, the functions $\psi$ and $\omega$ represent the two dynamical degrees of freedom of the gravitational field, in which $\psi$ accounts for the polarization mode $+$ while $\omega$ the polarization mode $\times$ \cite{thorne}. The function $\gamma$ plays the role of the gravitational energy of the system; it is connected to the C-energy \cite{stachel,thorne,goncalves}, more precisely $\gamma (\rho,u)$ gives the total energy per unit length enclosed within a cylinder of radius $\rho$ at the time $u$.

The dynamics of the spacetime is fully described by the coupled wave equations (\ref{eq2}) and (\ref{eq3}) for the functions $\psi(u,\rho)$ and $\omega(u,\rho)$. The Eqs. (\ref{eq4}) and (\ref{eq5}) determine the function $\gamma(u,\rho)$ from which, as we are going to see, the Bondi mass and the news functions can be defined. To evolve the spacetime it is necessary to specify the initial data expressed as,

\begin{equation}
\psi_0(\rho) \equiv \psi(\rho,u_0),\;\; \omega_0(\rho) \equiv \omega(\rho,u_0). \label{eq6}
\end{equation}

\noindent According to the characteristic scheme, the initial distributions $\psi_0(\rho)$ and $\omega_0(\rho)$ are free from any constraint relation, but must satisfy the requirements of regularity. Notice that if $\omega_0(\rho)=0$ implies that $\omega(u,\rho)=0$ in all subsequent instants.

The boundary conditions reflect the coordinate and regularity conditions for the spacetime. Performing a direct inspection of the field equations, the conditions of regularity and flatness of the metric near the origin $\rho=0$ impose that,

\begin{eqnarray}
&&\psi(\rho,u)=\mathrm{constant} + \mathcal{O}(\rho^2) \label{eq7} \\
\nonumber \\
&&\omega(\rho,u)=\mathcal{O}(\rho^2),\label{eq8}
\end{eqnarray}

\noindent The second boundary conditions are specified at the outer boundary or at the future null infinity, $\mathcal{J}^+$ ($\rho=\infty$). It can be shown \cite{stachel} that the asymptotic analysis of the wave equations (\ref{eq2}) and (\ref{eq3}) results in,

\begin{eqnarray}
&&\frac{\psi(\rho,u)}{\rho^{1/2}} = \mathcal{O}(\rho^{-1}),
\label{eq9} \\
\nonumber \\
&&\frac{\omega(\rho,u)}{\rho^{1/2}}=\beta(u) + \mathcal{O}(\rho^{-1}),\label{eq10}
\end{eqnarray}

\noindent where $\beta(u)$ is an arbitrary function. Notice $\psi$ decays in series of $\rho^{-1/2-n}$ ($n=0,1,2..$) instead of integers powers of the radial coordinate, and $\omega$ does not vanish at $\mathcal{J}^+$. It means that the spacetime is not asymptotically flat. 

We have followed the work of Stachel \cite{stachel} and presented the relevant physical aspects of cylindrical gravitational waves in the characteristic scheme. We start with the definition of the news functions that encompass the flow of information or the mass carried out by the gravitational radiation to infinity \cite{bondi_sachs}. Stachel \cite{stachel} had defined the news functions as the following asymptotic quantities,

\begin{eqnarray}
&&\frac{d c_1}{d u}\equiv \lim_{\rho \rightarrow \infty}\,\rho^{\frac{1}{2}}\psi_{,u}
\label{eq11} \\
\nonumber \\
&&\frac{d c_2}{d u}\equiv \lim_{\rho \rightarrow \infty}\,\rho^{\frac{1}{2}}\left(\frac{\mathrm{e}^{2\psi}\omega}{2\rho}\right)_{,u}.\label{eq12}
\end{eqnarray}

\noindent These two news functions are associated with each degree of freedom of the gravitational waves.

The definition of the cylindrical analog of the Bondi mass arises from the concept of the mass aspect in cylindrical symmetry. Thorne \cite{thorne} and Stachel \cite{stachel} arrived to the same result in identifying the function $\gamma(u,\rho)$ as the measure the amount of energy per unit of length enclosed by a cylinder of radius $\rho$ at the time $u$. Consequently, the Bondi mass in cylindrical symmetry, $M(u)$, is proportional to the asymptotic value of $\gamma(u,\rho)$ or,

\begin{equation}
M(u) = \frac{1}{2}\lim_{\rho \rightarrow \infty} \gamma. \label{eq13}
\end{equation}

\noindent This definition is valid in the case $\psi$ does not contain a static term $a \ln \rho +b$, otherwise we have to remove the infinite contribution according to $M(u) = \frac{1}{2}\lim_{\rho \rightarrow \infty} (\gamma - a^2 \ln \rho)$ since $\gamma_{\mathrm{static}}=a^2 \ln \rho + b$. As mentioned, the above definition agrees with the definition of the mass per unit of length obtained by Thorne \cite{thorne} that has followed a different approach with the introduction of the $C$-energy. Another useful expression to calculate the Bondi mass arises after the integration of Eq. (\ref{eq4}),

\begin{equation}
M(u) = \int_0^\infty\,\left(\rho \psi_{,\rho}^2 + \frac{\mathrm{e}^{4 \psi}}{4 \rho} \omega_{,\rho}^2\right) d \rho. \label{eq14}
\end{equation}

\noindent It shows the contribution of both degrees of freedom of the gravitational wave to the Bondi mass explicitly.

We now take into account the asymptotic of the field equation (\ref{eq5}) together with the definition of the Bondi mass (\ref{eq13}), we arrive at the Bondi formula in cylindrical symmetry,

\begin{equation}
\frac{d M}{d u}=-\left[\left(\frac{d c_1}{d u}\right)^2+\left(\frac{d c_2}{d u}\right)^2\right],\label{eq15}
\end{equation}

\noindent that states that the mass per unit of length always decreases if there are any news functions.

\vspace{0.2cm}
\section{The numerical scheme using the Galerkin-Collocation method}

We have adopted a spectral code based on the Galerkin-Collocation method \cite{gar_col} to integrate the wave equations (\ref{eq2}) and (\ref{eq3}). The central idea of any spectral method is to approximate the relevant fields $\psi$ and $\omega$ as appropriate series with respect to sets of certain basis functions. It will be convenient first to introduce a new radial coordinate $y$,

\begin{eqnarray}
\rho=y^2, \label{eq16}
\end{eqnarray}

\noindent followed by the new fields $\bar{\psi}$ and $\bar{\omega}$, respectively by \cite{clarke_dinverno},

\begin{eqnarray}
\bar{\psi}=y \psi \label{eq17} \\
\nonumber \\
\bar{\omega}=\frac{\omega}{y}. \label{eq18}
\end{eqnarray}

\noindent With the new radial variable the asymptotic expressions for $\bar{\psi}$ and $\bar{\omega}$ consist of powers of $1/y$ which is compatible with known analytical functions.

The spectral approximations for the metric functions $\bar{\psi}(u,y)$ and $\bar{\omega}(u,y)$ are,

\begin{eqnarray}
&& \bar{\psi}_a = \sum_{k=0}^{N_\psi}\,a_k(u) \Psi_k(y) \label{eq19} \\
\nonumber \\
&& \bar{\omega}_a = \sum_{k=0}^{N_\omega}\,b_k(u) \Phi_k(y), \label{eq20}
\end{eqnarray}

\noindent where $N_\psi$ and $N_\omega$ are the truncations orders, not necessarily equal, that dictate the number of unknow modes $a_j(u)$ and $b_k(u)$, respectively. According to the Galerkin method, the basis functions ${\Psi_j(y)}$ and ${\Phi_k(y)}$ satisfy the following boundary conditions: $\Psi_j = \mathcal{O}(y)$, $\Phi_k = \mathcal{O}(y^3)$, near the origin $y=0$, and $\Psi_j = \mathrm{constant} + \mathcal{O}(y^{-2})$, $\Phi_k = \mathrm{constant} + \mathcal{O}(y^{-1})$. In order to reproduce these conditions we have constructed basis functions as suitable combinations (see the Appendix) of the rational Chebyshev functions \cite{boyd},

\begin{equation}
TL_k(y) = T_k\left(x=\frac{y-L_0}{y+L_0}\right),\label{eq21}
\end{equation}

\noindent where $T_k(x)$ represents the standard Chebyshev polynomials of $k$-order and $L_0$ is the map parameter.

The next step is to substitute the spectral approximations (\ref{eq19}) and (\ref{eq20}) for the new fields into the wave equations (\ref{eq2}) and (\ref{eq3}) with the new radial variable $y$ to obtain the corresponding residual equations,

\begin{widetext}
\begin{eqnarray}
\mathrm{Res}_{\bar{\psi}}(u,y)&=& y\bar{\psi}_{a,uy} - \frac{{\rm e}^{\frac{4\bar{\psi}_a}{y}}}{2y} (\bar{y\omega}_a)_{,y} \bar{\omega}_{a,u} - \frac{1}{4}\left[y\left(\frac{\bar{\psi}_a}{y}\right)_{,y}\right]_{,y} + \frac{{\rm e}^{\frac{4\bar{\psi}_a}{y}}}{8y^3} (y\bar{\omega}_a)_{,y}^2  \label{eq22}\\
\nonumber \\
\mathrm{Res}_{\bar{\omega}}(u,y)&=&y\bar{\omega}_{a,uy} + \frac{2}{y} (y\bar{\omega}_{a})_{,y}\bar{\psi}_{a,u} + 2 y \left(\frac{\bar{\psi}_a}{y}\right)_{,y} \bar{\omega}_{a,u} - \frac{y^2}{4}\left[\frac{(y\bar{\omega}_a)_{,y}}{y^3}\right]_{,y} - \frac{1}{y}(y\bar{\omega}_{a})_{,y}\left(\frac{\bar{\psi}_a}{y}\right)_{,y}. \label{eq23}
\end{eqnarray}
\end{widetext}

\noindent In general the residuals $\mathrm{Res}_{\bar{\psi}}(u,y)$ and $\mathrm{Res}_{\bar{\omega}}(u,y)$ do not vanish since $\bar{\psi}_a$ and $\bar{\omega}_a$ are approximations to the exact $\bar{\psi}$ and $\bar{\omega}$. According to the Collocation method these residual equations vanish at the collocation or grid points. Schematically we have,

\begin{eqnarray}
\mathrm{Res}_{\bar{\psi}}(u,y_k)&=& 0,\;\; k=0,1,..,N_{\psi}  \label{eq24}\\
\nonumber \\
\mathrm{Res}_{\bar{\omega}}(u,y_k)&=& 0,\;\; k=0,1,..N_{\omega}. \label{eq25}
\end{eqnarray}

\noindent Here $y_k$ denotes the collocation points in the physical domain that are calculated from the Chebyshev-Gauss points $x_k$,

\begin{equation}
x_k = \cos\left(\frac{(2k+1)\pi}{2N+2}\right), \label{eq26}
\end{equation}

\noindent with $k=0,1,..,N$ and $N=N_\psi,N_\omega$ using the algebraic map $y_k = L_0 (1+x_k)/(1-x_k)$.

We have approximated the field equations into a set of ordinary differential equations written in the following matricial form,

\begin{eqnarray}
\textbf{M} \begin{pmatrix} \partial \bar{\psi}_k \\  \\  \partial \bar{\omega}_j \end{pmatrix} = \mathbf{B} \label{eq27}
\end{eqnarray}

\noindent for all $k=0,1,..,N_\psi$ and $j=0,1,.,N_\omega$. In the above expression we have,

\begin{eqnarray}
\partial \bar{\psi}_k(u) \equiv \left(\frac{\partial \bar{\psi}_a}{\partial u}\right)_k = \sum_{i=0}^{N_\psi}\,a_{i,u}(u) \Psi_i(y_k) \label{eq28}\\
\nonumber \\
\partial \bar{\omega}_j(u) \equiv \left(\frac{\partial \bar{\omega}_a}{\partial u}\right)_j = \sum_{i=0}^{N_\psi}\,b_{i,u}(u) \Phi_i(y_k) \label{eq29}
\end{eqnarray}

\noindent where $\partial \bar{\psi}_k(u)$ and $\partial \bar{\omega}_j(u)$ are the values of the derivatives of $\bar{\psi}$ and $\bar{\omega}$ with respect to $u$ at the collocation points. Note that these values are related to the time derivatives of the unknown modes $a_{k,u}(u),b_{j,u}(u)$. The matrices $\textbf{M}$ and $\textbf{B}$ depend on the unknown modes $a_k(u),b_j(u)$ as well the values of $\bar{\psi}$ at the collocation points, or

\begin{eqnarray}
\bar{\psi}_k(u) \equiv \bar{\psi}_a(u,y_k) = \sum_{i=0}^{N_\psi}\,a_i(u) \Psi_i(y_k), \label{eq30}
\end{eqnarray}

\noindent that provides a set of relations between the values and the unknown modes. The integration processes as follows: starting from the initial modes $a_k(u_0),b_k(u_0)$ we can determine the initial values $\bar{\psi}_k(u_0)$ as well the initial matrices $\textbf{M},\textbf{B}$. The dynamical system gives the initial values $\partial \bar{\psi}_k(u_0),\partial \bar{\omega}_j(u_0)$ that allows to determine $a_{k,u}(u_0),b_{k,u}(u_0)$, and as a consequence, the modes at the next time step repeating the whole process. We have used a fourth-order Runge-Kutta integrator in all cases.

\section{Numerical Tests}

\subsection{Einstein-Rosen waves: testing the code}

The Einstein-Rosen waves \cite{ER} represent the exact non-static solution of the field equations when $\omega=0$. The field equation (\ref{eq2}) becomes a free wave equation for $\psi$ in cylindrical coordinates. For the sake of convenience we present a particular form of the solution obtained by Weber and Wheeler \cite{weber_wheeler} to test our code. Using the variables $(u,y)$ the Weber-Wheeler solution is expressed as,

\begin{widetext}
\begin{eqnarray}
\psi_{\mathrm{exact}}(u,y)=A_0\sqrt {{\frac {\sqrt {a^2+y^4+(u+y^2)^2\,[2a^2-2y^4+(u+y^2)^2]}+
a^2-u^2-2\,uy^2}{a^2+y^4+(u+y^2)^2\,[2a^2-2y^4+(u+y^2)^2]}}} \label{eq31}
\end{eqnarray}
\end{widetext}

\noindent where $A_0$ and $a$ are constants identified as the amplitude and the width of the wave, respectively. Physically, $\psi_{\mathrm{exact}}(u,y)$ represents an ingoing gravitational wave with polarization mode $+$ that hits the axis of symmetry and rebound back to infinity. One can obtain the exact Bondi mass and the news functions after a straightforward calculation with the solution (\ref{eq31}).

We have tested the spectral code by comparing the exact solution (\ref{eq31}) with the numerical solution obtained with the initial data $\psi_0(y)=\psi_{\mathrm{exact}}(u_0,y)$ and $\omega_0(y)=0$, where in this case the field equations yield $\omega(u,y) = 0$. We have considered two numerical tests. The first is to compare the exact and approximate Bondi masses by evaluating the deviation $\delta M$ given by,

\begin{equation}
\delta M = \left[\frac{1}{\Delta u}\,\int_0^{\Delta u}\,(M_{\mathrm{exact}}-M(u))^2 du \right]^{\frac{1}{2}},\label{eq32}
\end{equation}

\noindent where we have assumed that $a=2,\Delta u=7.0$ and $A_0=1.0$. We have calculated the deviation $\delta M$ for the truncation orders $N_\psi=20,30,..,80$ and, as expected, the graph of Fig. 1 shows the exponential decay of $\delta M$.

\begin{figure}[h]
\begin{center}
\includegraphics*[scale=0.27]{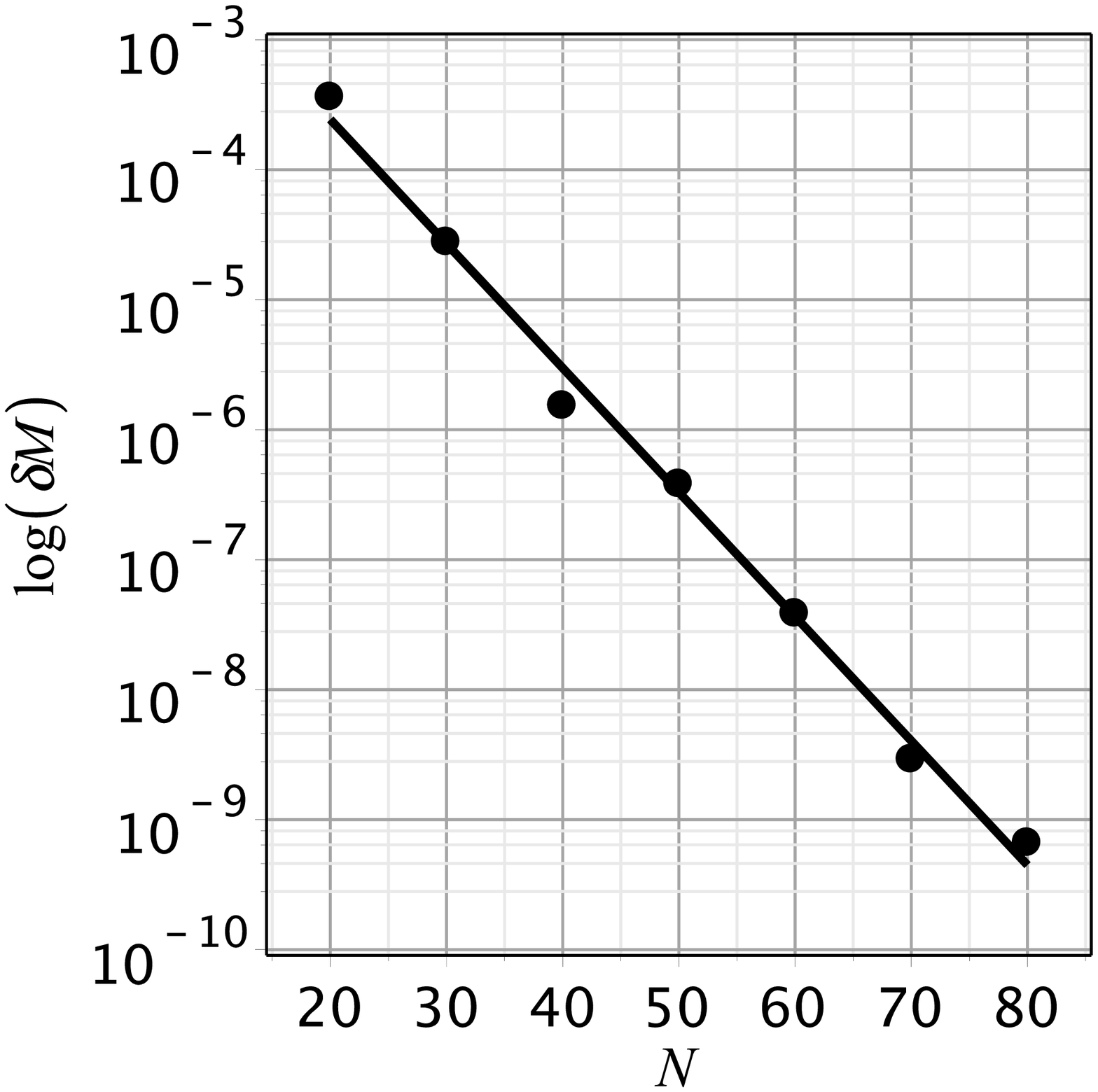}
\includegraphics*[scale=0.27]{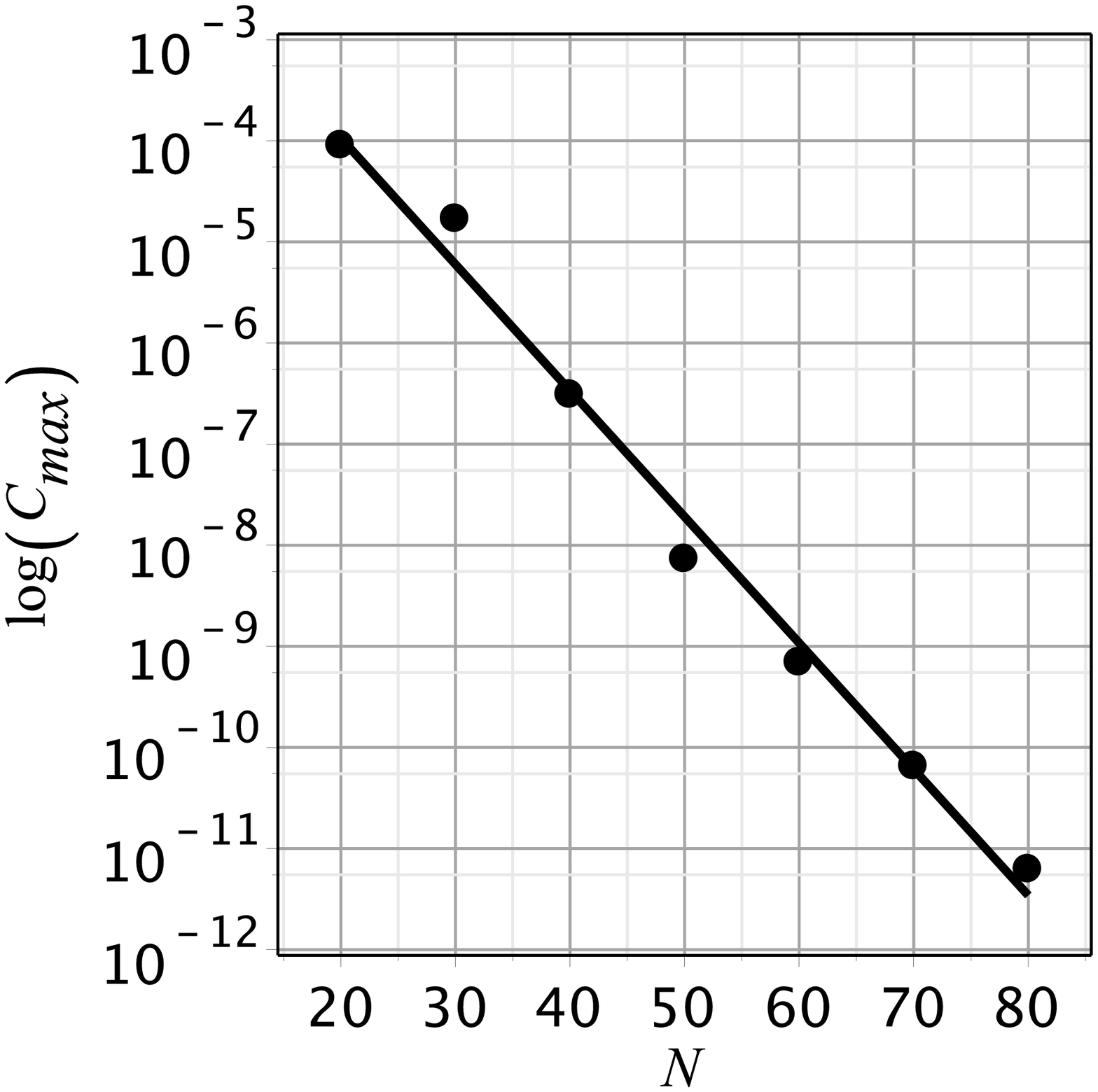}
\end{center}
{\renewcommand{\baselinestretch}{1}
\caption{Exponential decays of $\delta M$ and $C_{max}$ shown by the upper and lower panels.}}
\end{figure}

The second test consists in the verification of the global energy conservation provided by the Bondi formula expressed as \cite{winicour},

{\small{
\begin{equation}
\mathcal{C}(u) = \frac{M(u)-M_0}{M_0}+ \frac{1}{M_0}\,\int_{u_0}^u\,\left[\left(\frac{d c_1}{d u}\right)^2+\left(\frac{d c_2}{d u}\right)^2\right] du,\label{eq33}
\end{equation}}}

\noindent where $M_0=M(u_0)$ is the initial Bondi mass and $dc_2/du=0$ for the case of a polarized wave. Whereas exact solution (\ref{eq31}) yields $\mathcal{C}(u)=0$, any deviation indicates the error of the numerical solution. We have proceeded using truncation orders $N_\psi=20,30,..,80$, evolved the field equations until $u=7.0$ and selected the maximum deviation, $C_{max}$, for each truncation order. The result is presented in Fig. 1 with the exponential decay of the maximum deviation $C_{max}$.

\subsection{Nonlinear evolution}

We present now some results of the nonlinear evolution characterized when the initial function $\bar{\omega}(u_0,y) \neq 0$, and both modes of polarization are present. We have chosen an initial data of compact support representing an initial pulse of the gravitational wave.

\begin{eqnarray}
\bar{\psi}_0(y) = A_0 y^2\mathrm{e}^{-(y-y_1)^2/\sigma_1^2} \label{eq34} \\
\nonumber \\
\bar{\omega}_0(y) = \frac{B_0  y^3}{1+y^2}\mathrm{e}^{-(y-y_2)^2/\sigma_2^2} \label{eq35}
\end{eqnarray}

\noindent where $A_0,B_0$ represent the initial amplitudes of the wave modes, the constants $y_1,y_2$ denote the position and $\sigma_1,\sigma_2$ the widths of the waves. From these expression we can calculate the initial modes $a_j(u_0)$ and $b_k(u_0)$ to evolve the dynamical equations (\ref{eq28}) and (\ref{eq29}). We have fixed $y_1=1.0,y_2=1/3$ and $\sigma_1=1,\sigma_2=2$.

We have provided two convergence tests for the nonlinear evolution of cylindrical waves. The first we have borrowed from Piran et al. \cite{piran} that consists of comparing the values of $\gamma(u,y)$ evaluated from the spatial and time integrations of Eqs. (\ref{eq4}) and (\ref{eq5}), respectively. We have chosen $y=4.0$ and $u=0.14$ taking into account increasing truncation orders $N_\psi$ and $N_\omega$. We have presented the resulting showing the convergence tests for $\gamma$ in both graphs of Fig. 2. It is worth of mentioning that the similar convergence is observed for other values of $y$ and $u$.

\begin{figure}[h]
\begin{center}
\includegraphics*[width=5.8cm,height=4.7cm]{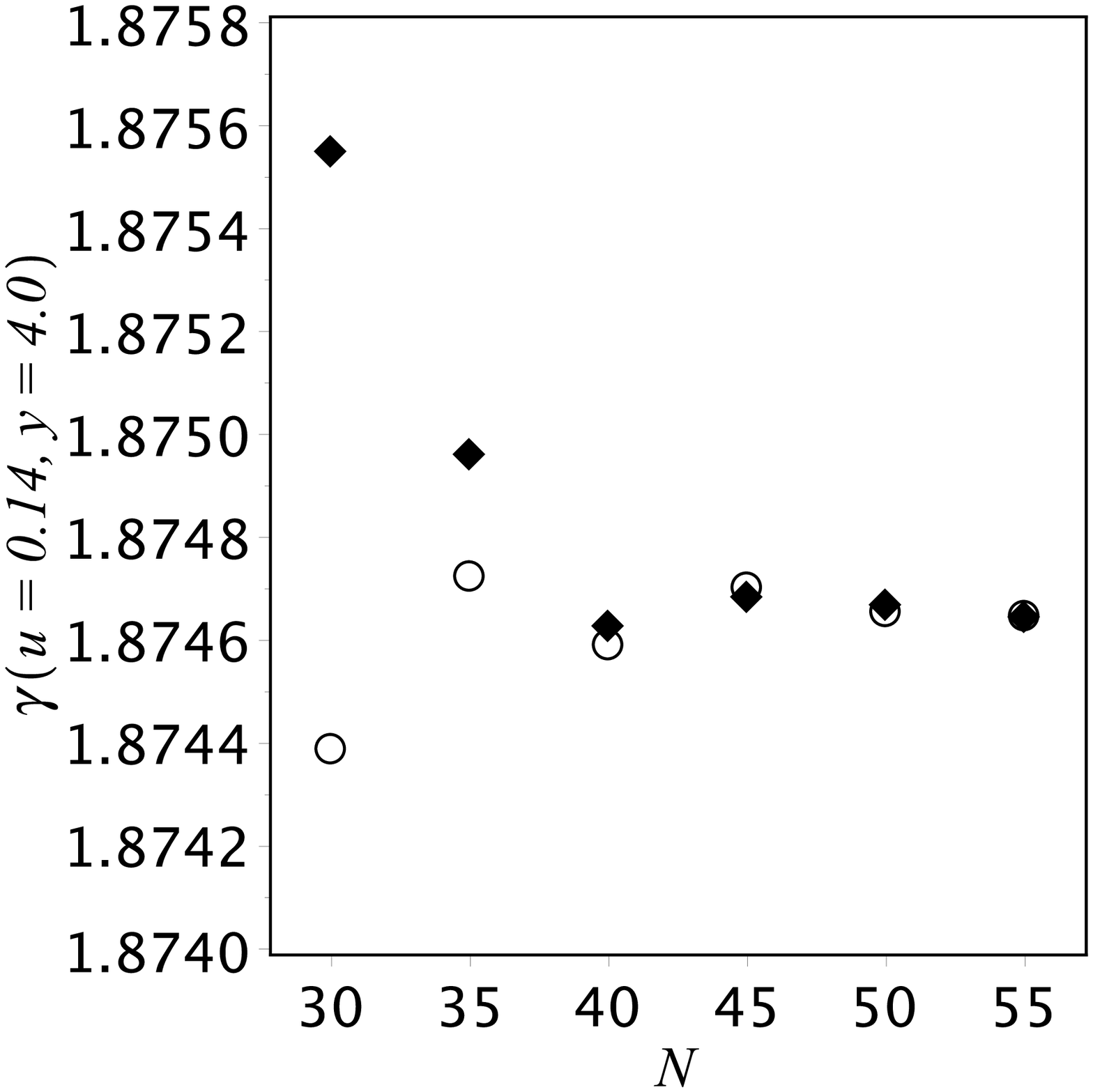}
\includegraphics*[width=5.8cm,height=4.7cm]{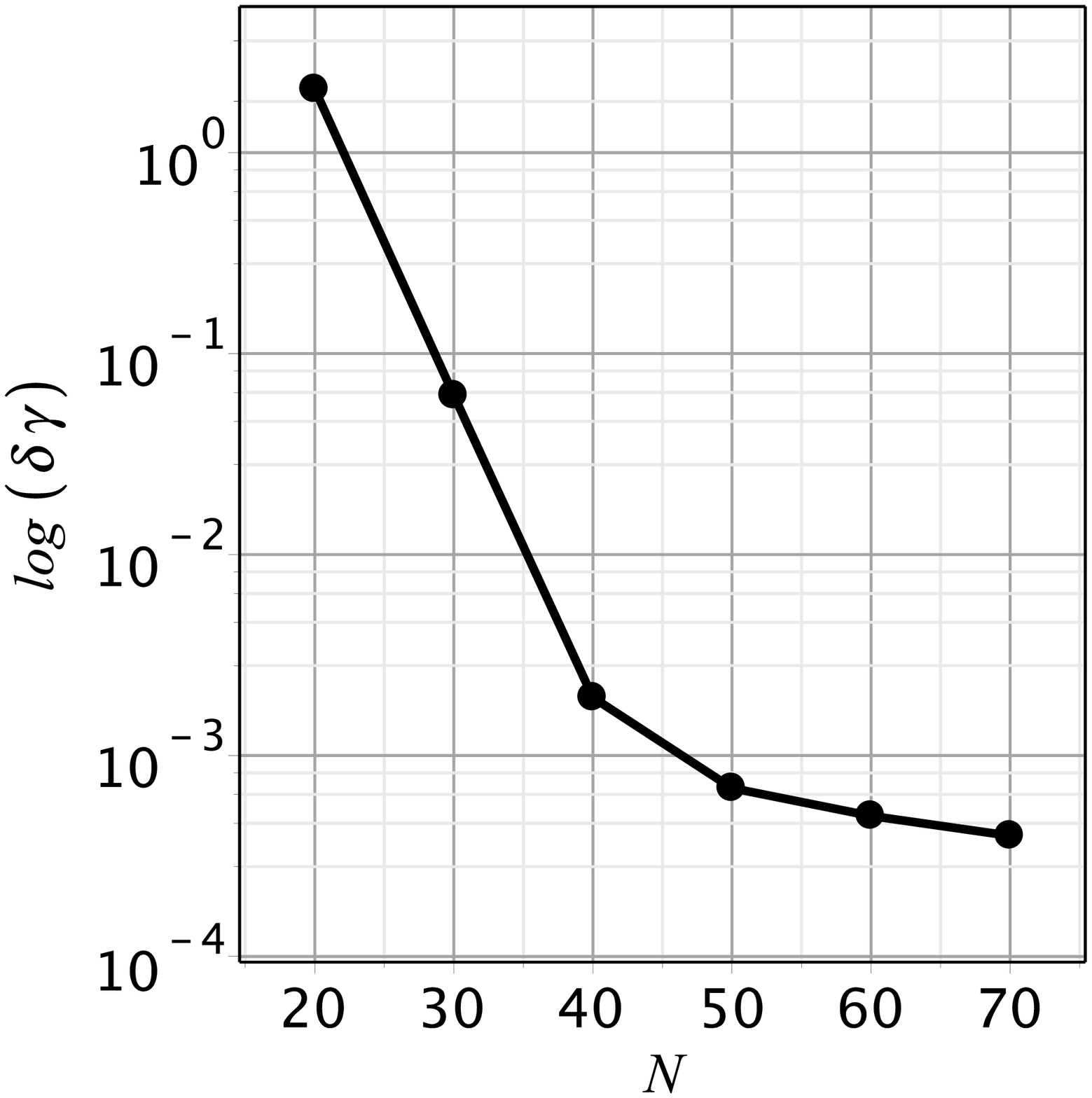}
\end{center}
\vspace{-0.5cm}
\caption{Convergence of $\gamma$ evaluated at $y=4.0$ and $u=0.14$ after integrating the Eqs. (\ref{eq4}) and (\ref{eq5}) (circles and diamonds respectively). In the lower plot the decay of the difference between these values for each truncation order. We have set $A_0=1.0$ and $B_0=0.7$ for the initial data (\ref{eq34}) and (\ref{eq34}). Here $\delta \gamma=|\gamma^{(u)}-\gamma^{(\rho)}|/\gamma^{(\rho)} \times 100$, where $\gamma^{(\rho)},\gamma^{(u)}$ result from the integration of Eqs. (\ref{eq4}) and (\ref{eq5}), respectively.}
\end{figure}


The second test is the verification of the global energy conservation provided the Bondi formula expressed by Eq. (\ref{eq33}). In the nonlinear case, the cylindrical waves are unpolarized meaning that both news functions are present. In Fig. 3 we have illustrated the qualitative agreement between the decay of the Bondi mass with the amount of mass carried out by the gravitational waves. This last quantity is the integral of the integral in time of the rhs of Eq. (\ref{eq15}). Fig. 4 exhibits the exponential decay of $C_{max}$ until reaching to its saturation value for $N=N_\psi=N_\omega \geq 50$.

\begin{figure}[h]
\begin{center}
\includegraphics*[scale=0.24]{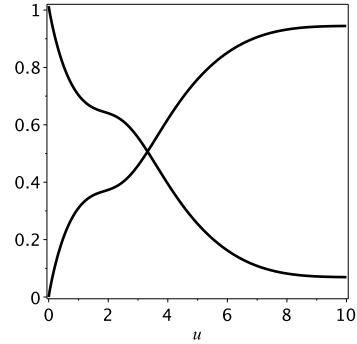}
\end{center}
\vspace{-0.5cm}
\caption{Decay of the Bondi mass together with the integral of the news functions for the evolution of nonlinear waves starting from the initial data (\ref{eq34}) and (\ref{eq35}) with $A_0=1.0$ and $B_0=0.7$. In this illustration we have considered $N=N_\psi=N_\omega=40$.}
\end{figure}

\begin{figure}[h]
\begin{center}
\includegraphics*[scale=0.27]{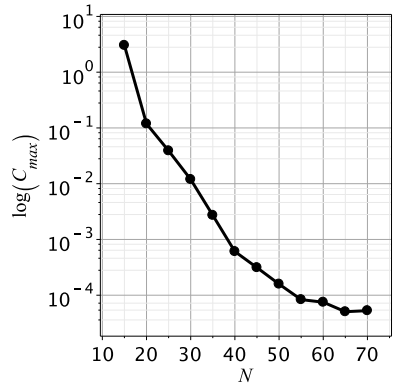}
\end{center}
\vspace{-0.5cm}
{\renewcommand{\baselinestretch}{1}
\caption{Exponential decay of the maximum deviation of the energy conservation for each truncation order $N=N_\psi=N_\omega$. After $N \geq 50$ the maximum error approaches to its saturation value.}}
\end{figure}

\section{Physical aspects}

\subsection{Nonlinear interaction between the polarization modes $+$ and $\times$}

We discuss here the consequences of the nonlinear interaction between the gravitational waves of distinct polarization modes in the realm of the characteristic scheme. To this aim we have looked closely to the process of mass extraction described by the Bondi formula (\ref{eq15}). Accordingly, the news functions associated to both polarization modes dictate the amount of Bondi mass carried away. For the sake of convenience, it will be useful to define the following quantities,

\begin{eqnarray}
I_1(u) &=& \int_{u_0}^u\,\left(\frac{dc_1}{du}\right)^2 du \label{eq36}\\
\nonumber \\
I_2(u) &=& \int_{u_0}^u\,\left(\frac{dc_2}{du}\right)^2 du, \label{eq37}
\end{eqnarray}

\noindent that are the amount of mass extracted by the gravitational waves modes $\psi$ and $\omega$, respectively.


We have already mentioned that if $\bar{\omega}_0(y)=0$ then $\bar{\omega}_0(u,y)=0$ at any instant $u > u_0$ as a direct consequence of the field equations. Thus, in this case $I_2(u)=0$ for all $u>u_0$. However, we are interested in the behavior of $I_1(u)$ and $I_2(u)$ generated with the initial data $\bar{\psi}_0(y)=0$ and $\bar{\omega}_0(y) \neq 0$ ($A_0=0$ and $B_0 \neq 0$ in Eqs. (\ref{eq34}) and (\ref{eq35})). Due to the nonlinear interaction between the distinct wave modes, $\bar{\psi}(u,y)$ will be excited and both wave modes will act in the mass extraction process.


The numerical simulations consist in evolving cylindrical waves with the initial data (\ref{eq34}) and (\ref{eq35}) with $A_0=0$ and $B_0$ as a free parameter. We have expressed the results in Fig. 5 by a sequence of joint plots of $I_1(u)$ and $I_2(u)$. The first aspect to be noticed is that $I_1(u)$ saturates more quickly than $I_2(u)$ irrespective to the value of $B_0$. In general we should expect $I_2(u) > I_1(u)$ throughout all evolution since $A_0=0$, but this is true for those values of $B_0$ smaller than certain critical value, $B_0^{(\mathrm{crit)}} \approx 2.0114$, from which the asymptotic values of both $I_1(u)$ and $I_2(u)$ are approximately equal. For values $B_0 > B_0^{(\mathrm{crit)}}$ it follows $I_1(u) > I_2(u)$ for all $u > u_0$ signalizing the dominance of the mode $\psi$ in extracting mass. In other words, the growth of the mode $\psi$ is such that it starts to be the dominant channel from which mass is carried away.  We call this feature as the \textit{enhancement effect} resulting from the nonlinear interaction between the modes $\psi$ and $\omega$.

\begin{figure}[h]
\begin{center}
\includegraphics*[width=7.5cm,height=10cm]{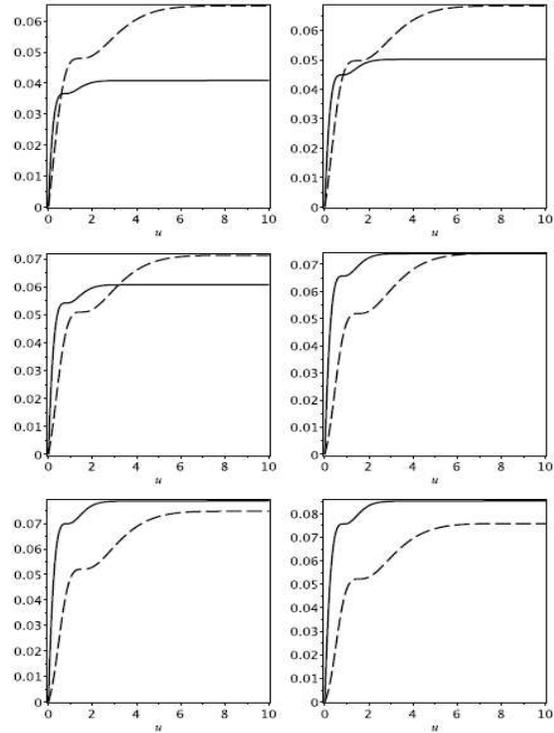}
\end{center}
\caption{Behavior of $I_1(u)$ (continuous line) and $I_2(u)$ (dashed line) starting with $\bar{\psi}(0,y)=0$ and $\bar{\omega}(0,y) \neq 0$, or $A_0=0$ and $B_0 \neq 0$, respectively in Eqs. (\ref{eq34}) and (\ref{eq35}). The graphs correspond to $B_0=1.7,1.8,1.9,2.011,2.05,2.1$ from left to right and up to down. Notice the enhancement effect of the mode $+$ due to the nonlinear interaction with the mode $\times$. We have defined the critical value $B_0^{\mathrm{(crit)}} \approx 2.011$ such that the asymptotic values of $I_1(u)$ and $I_2(u)$ are approximately equal.}
\end{figure}

\subsection{About the matter fields in cylindrical symmetry}

There are several works dealing with the cylindrical collapse of matter fields such as dust \cite{nakao}, shells of dust and perfect fluids \cite{apostolatos,echeverria,nakao2}, null fluids \cite{morgan,letelier,nakao3}, dissipative fluids \cite{fatima,herrera} and scalar fields \cite{wang,nolan}. The main motivations are concerning the cosmic censorship hypothesis, the emission of gravitational waves during the collapse and the critical phenomena. In most of them, the line element is not written in the hyperbolic canonical form we are adopting, and also the additional degree of freedom represented by the function $\omega$ is absent. In the present discussion, we are going to consider the line element in the hyperbolical canonical form (see Eq. \ref{eq1}). In order to include any matter field, the following condition must be satisfied by the corresponding energy-momentum tensor,

\begin{equation}
T_u\,^u + T_\rho\,^\rho = 0, \label{eq38}
\end{equation}

\noindent Then, null fluids and massless scalar fields, for instance, satisfy the above condition. However, these matter fields do not couple with the gravitational potentials $\psi$ or $\omega$. Only equations (\ref{eq4}) and (\ref{eq5}) are altered by these matter fields adding contributions to the Bondi mass and news functions. In particular, considering $\omega=0$ and introducing a massless scalar field, $\varphi(u,\rho)$, the field equations are equivalent to a pure gravitational wave after the change $\psi \rightarrow \psi + \varphi/\sqrt{2}$.



Electromagnetic fields satisfy the condition (\ref{eq38}) and couple with the gravitational potentials $\bar{\psi}$ and $\bar{\omega}$. It means that electromagnetic waves can generate gravitational radiation and the other way round. A similar conversion occurs in the scattering of electromagnetic waves by a black hole where gravitational radiation emerges as after been disturbed by a package of electromagnetic waves \cite{chandra_book}. The introduction of electromagnetic fields in cylindrical spacetimes has been discussed by Melvin \cite{melvin} with the cylindrical electromagnetic universes (CEU) and the presentation of the most general static configuration for such spacetimes.

We consider briefly here the dynamic of electromagnetic fields in cylindrical spacetimes. Following Thorne \cite{thorne} we have set $\omega=0$ and assumed that the potential vector components $A_\mu = (0,0,A_2,A_3)$ depend on the coordinates $u$ and $\rho$ (or $y$) such that the resulting electric and magnetic fields lie on the plane $z \phi$. The field equations read as,

\begin{widetext}
\begin{eqnarray}
&& y\bar{\psi}_{,uy} - \frac{1}{4}\left[y\left(\frac{\bar{\psi}}{y}\right)_{,y}\right]_{,y} + y\mathrm{e}^{-\frac{2\bar{\psi}}{y}} \left(\frac{\bar{A}_2}{y}\right)_{,y}\left[\bar{A}_{2,u} - \frac{1}{4}\left(\frac{\bar{A}_2}{y}\right)_{,y} \right]  -  y\mathrm{e}^{\frac{2\bar{\psi}}{y}} (y\bar{A}_3)_{,y}\left[\bar{A}_{3,u} - \frac{\left(y\bar{A}_3\right)_{,y}}{4y^2}\right] = 0 \label{eq39} \\
\nonumber \\
&& y\bar{A}_{2,uy} - \frac{1}{4}\left[y\left(\frac{\bar{A}_2}{y}\right)_{,y}\right]_{,y} - y \left(\frac{\bar{A}_2}{y}\right)_{,y}\bar{\psi}_{,u} - y \left(\frac{\bar{\psi}}{y}\right)_{,y}\bar{A}_{2,u} + \frac{y}{2} \left(\frac{\bar{\psi}}{y}\right)_{,y} \left(\frac{\bar{A}_2}{y}\right)_{,y}  = 0 \\
\nonumber \label{eq40} \\
&& y\bar{A}_{3,uy} - \frac{y^2}{4}\left[\frac{(y \bar{A}_3)_{,y}}{y^3}\right]_{,y} + \frac{1}{y} \left(y\bar{A}_3\right)_{,y}\bar{\psi}_{,u} + y\left(\frac{\bar{\psi}}{y}\right)_{,y}\bar{A}_{3,u} - \frac{1}{2y} \left(\frac{\bar{\psi}}{y}\right)_{,y} \left(y\bar{A}_3\right)_{,y}  = 0. \label{eq41}
\end{eqnarray}
\end{widetext}



\noindent In the above equations $\bar{A}_2=yA_3$ and $\bar{A}_3=A_3/y$. We have adapted the spectral code for the evolution of pure gravitational waves to the present case (now  $\bar{\psi}_0(y)=\bar{\psi}(u_0,y), \bar{A}_{20}(y)=\bar{A}_2(u_0,y)$ and $\bar{A}_{30}(y)=\bar{A}_3(u_0,y)$ constitute the initial data). The corresponding spectral approximation for the potentials $\bar{A}_2(u,y)$ and $\bar{A}_3(u,y)$ use the same basis functions for the gravitational potentials $\bar{\psi}$ and $\bar{\omega}$, respectively. It means that the potentials $\bar{\psi}(u,y), \bar{A}_2(u,y)$ and $\bar{\omega}(u,y), \bar{A}_3(u,y)$ obey the same boundary conditions. Moreover, the similarity between these potentials can be inferred by inspecting the corresponding field equations.

Before exhibiting some numerical results, we list the amended expressions for the Bondi mass and the Bondi formula modified by the introduction of the electromagnetic field. Expressing the Bondi mass as an integral analogous to Eq. (\ref{eq14}) and taking into account the functions $\bar{\psi},\bar{A}_2,\bar{A}_3$ and the new coordinate $y$, we obtain,

\begin{eqnarray}
M(u) = &&\frac{1}{2} \int_0^\infty\,\Bigg[y\left(\frac{\bar{\psi}}{y}\right)^2_{,y} + y \mathrm{e}^{-\frac{2\bar{\psi}}{y}}\left(\frac{\bar{A}_2}{y}\right)^2_{,y} \nonumber\\
 && + \frac{\mathrm{e}^{\frac{2\bar{\psi}}{y}}}{y^3}\left(y\bar{A}_3\right)^2_{,y}\Bigg] dy. \label{eq42}
\end{eqnarray}

\noindent The second and third terms of the integral are contributions of the electromagnetic field to the Bondi mass of the system. The Bondi formula is straightforwardly derived from the field equations for the function $\gamma(u,y)$ (not presented here),

\begin{equation}
\frac{d M(u)}{du} = -\left[\left(\frac{dc_1}{du}\right)^2 + \left(\frac{dc_2}{du}\right)_{\mathrm{e}}^2 + \left(\frac{dc_3}{du}\right)_{\mathrm{e}}^2\right], \label{eq43}
\end{equation}

\noindent where,

\begin{eqnarray}
\left(\frac{dc_2}{du}\right)_{\mathrm{e}} = \lim_{y \rightarrow \infty}\, \mathrm{e}^{-\frac{2\bar{\psi}}{y}}\bar{A}_{2,u} \label{eq44}\\
\nonumber \\
\left(\frac{dc_3}{du}\right)_{\mathrm{e}} = \lim_{y \rightarrow \infty}\, \mathrm{e}^{\frac{2\bar{\psi}}{y}}\bar{A}_{3,u}, \label{eq45}
\end{eqnarray}
	
\noindent are the news functions associated to the electromagnetic potentials $\bar{A}_2$ and $\bar{A}_3$, respectively.


We have explored some consequences of the interaction of electromagnetic and gravitational waves. Following the case of unpolarized waves, we have focused on the behavior of $I_1(u)$ resulting from the initial gravitational potential distribution $\bar{\psi}_0=\bar{\psi}(u_0,y)=0$, while $\bar{A}_2(u_0,y),\bar{A}_3(u_0,y)$ are given by the initial data functions (\ref{eq34}) and (\ref{eq35}), respectively. It is necessary to add the corresponding quantities that measure the amount of mass carried out by electromagnetic waves, $I^{(e)}_2(u)$ and $I^{(e)}_3(u)$, defined by,

\begin{eqnarray}
I^{(e)}_2(u) &=& \int_{u_0}^u\,\left(\frac{dc_2}{du}\right)_e^2 du \label{eq46}\\
\nonumber \\
I^{(e)}_3(u) &=& \int_{u_0}^u\,\left(\frac{dc_3}{du}\right)_e^2 du. \label{eq47}
\end{eqnarray}


\noindent We have evolved the field equations (\ref{eq39}) - (\ref{eq41}) such that $N_\psi=N_2=N_3=60$, where $N_{2,3}$ are the truncation orders of the spectral expansions of the electromagnetic potentials (not shown here).

\begin{figure}[h]
\begin{center}
\includegraphics*[width=7.5cm,height=10cm]{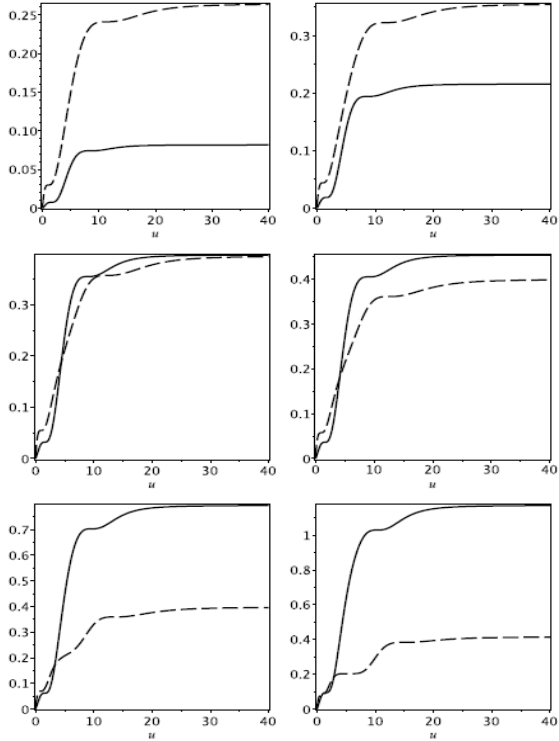}
\end{center}
\caption{Behavior of $I_1(u)$ (continuous line) and $I^{(e)}_2(u)$ (dashed line) in which $\bar{\psi}(0,y),\bar{A}_3(0,y)=0$ and $\bar{A}_2(0,y) \neq 0$ or $A_0 \neq 0$ in Eq. (\ref{eq34}). The graphs correspond to $A_0=0.70,0.90,1.06,1.10,1.30,1.50$ from left to right and up to down. The enhancement of the gravitational mode is effective here triggered by the interaction with the electromagnetic field.}
\end{figure}

The results are exhibited as sequences of plots of $I_1(u),I^{(e)}_2(u)$ and $I^{(e)}_3(u)$. Besides setting the initial gravitational potential null, we have considered the following cases: (i) $\bar{A}_2(u_0,y)\neq 0$, $\bar{A}_3(u_0,y)=0$ (Fig. 6), (ii) $\bar{A}_2(u_0,y)=0$, $\bar{A}_3(u_0,y)\neq 0$ (Fig. 7) and (iii) $\bar{A}_2(u_0,y),\bar{A}_3(u_0,y) \neq 0$ (Fig. 8). In the cases (i) and (ii) we notived that if one of the potentials vanishes initially, it remains null for all $u>u_0$. For this reason, there are two curves in Figs. 6 and 7, where the continuously line represents $I_1(u)$, and the dashed line $I^{(e)}_2(u)$ and $I^{(e)}_3(u)$, respectively.

A close inspection of Figs. 6 and 7 reveal that there is always an initial critical amplitude such that final value of $I_1(u)$ is greater than the corresponding values of $I^{(e)}_2(u)$ or $I^{(e)}_3(u)$. The enhancement effect is now due to the nonlinear interaction between electromagnetic and gravitational waves.

The last case is shown in Fig. 8 has the contribution of both electromagnetic fields with the behavior of $I^{(e)}_2(u)$ (dashed line) and $I^{(e)}_3(u)$ (dash-point line) together with $I_1(u)$ (continuous line). By changing both initial amplitudes $A_0$ and $B_0$ we noticed the reproduction of the enhancement effect. The critical amplitude depends on the initial amplitudes of the electromagnetic potentials. For the results shwon in Fig. 8 we have fixed $A_0=0.5$ and changed $B_0$.

\begin{figure}[h]
\begin{center}
\includegraphics*[width=7.5cm,height=10cm]{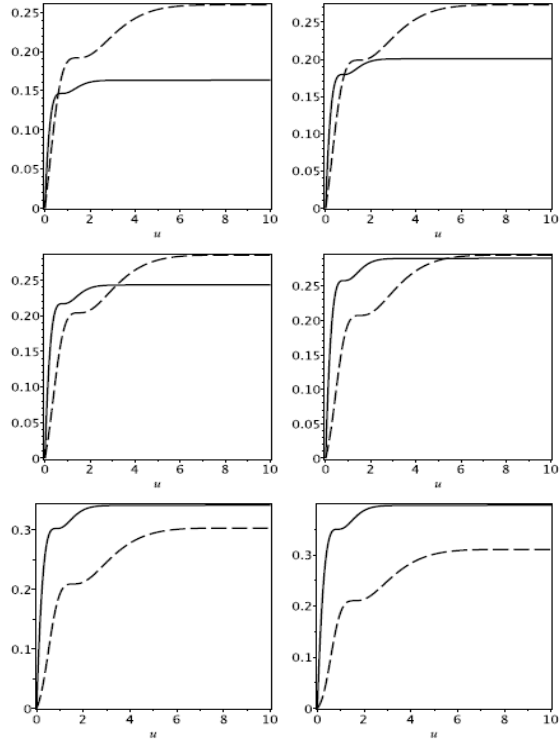}
\end{center}
\caption{Behavior of $I_1(u)$ (continuous line) and $I^{(e)}_3(u)$ (dashed line) in which $\bar{\psi}(0,y),\bar{A}_2(0,y)=0$ and $\bar{A}_3(0,y) \neq 0$ or $B_0 \neq 0$ in Eq. (\ref{eq35}). The graphs correspond to $A_0=0.5$ and $B_0=1.70,1.80,1.90,2.0,2.10,2.20$ from left to right and up to down. Again the enhacement effect takes place.}
\end{figure}

\begin{figure}[h]
\begin{center}
\includegraphics*[width=7.5cm,height=10cm]{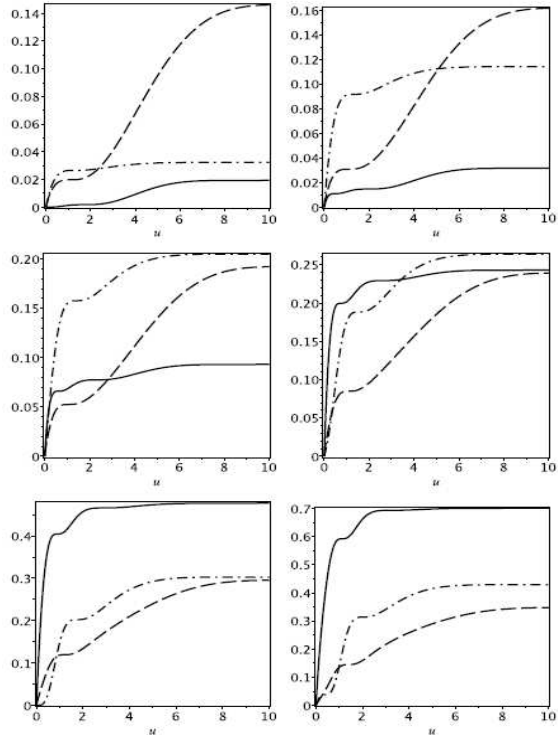}
\end{center}
{\renewcommand{\baselinestretch}{1}
\caption{Behavior of $I_1(u)$ (continuous line), $I^{(e)}_2(u)$ (dashed line) and $I^{(e)}_3(u)$ (dash-point line) in which $\bar{\psi}(0,y)=0$ and $\bar{A}_2(0,y),\bar{A}_3(0,y) \neq 0$ or $A_0,B_0 \neq 0$ in Eqs. (\ref{eq34}) and (\ref{eq35}) respectively. The graphs correspond to fixed $A_0=0.5$ and $A_0=0.5,1.0,1.5,2.0,2.5,3.0$ from left to right and up to down. The enhancement of the gravitational mode is again effective here but the critical amplitude depends on particular combinations of $A_0$ and $B_0$.}}
\end{figure}

\section{Conclusions}


In this paper, we have investigated the numerical evolution of general cylindrical gravitational waves using a code based on the Galerkin-Collocation method. We have established for the first time the numerical realization of the characteristic scheme adapted to cylindrical symmetry due to Stachel \cite{stachel}. Relevant quantities like the Bondi mass, news functions, and the Bondi formula were presented and studied numerically. Therefore, we have exhibited the decay of the Bondi mass due to the action of the news functions associated with the polarization modes of the gravitational waves.


Rigorously speaking, cylindrical symmetry is not of astrophysical interest, but it is the simplest axisymmetric spacetime in which the gravitational waves possess both polarization modes. Therefore, it is a valid theoretical arena for exploring the nonlinear interaction of these polarization modes. In this direction, Piran and Stark have shown the equivalent gravitational effect of the Faraday rotation. We have presented the enhancement effect that consists in the preferred increase of the mode $+$ even if this mode vanishes initially. We have noticed that by increasing the initial amplitude of the mode $\omega$ the amount of mass extracted by the mode $\psi$ increases such that above to certain critical value it becomes greater than the counterpart due to the mode $\omega$.


We have made a brief discussion of the role of matter fields in cylindrical symmetry focusing on the hyperbolical canonical form of the line element. In this case the most natural matter field that can interact directly with the gravitational waves is the electromagnetic field. Thus, we have extended the spectral code to integrate the Maxwell-Einstein equations that describe the electromagnetic universes \cite{melvin,thorne} in the scheme of characteristics. In this case, we have considered only one gravitational potential, $\psi$, together with the potential vector components compatible with the cylindrical symmetry. Therefore, the expressions for the Bondi mass and the news functions have now the signature of the electromagnetic field. We have performed some numerical investigation and shown the same effect of enhancement with the initially vanishing gravitational potential, $\psi(u_0,\rho)=0$. Due to the nonlinear interaction with electromagnetic potentials, the gravitational waves can be excited such that they start to extract most of the mass.


We would like to point out two possible lines of investigation. The first is the extension of the code to evolve electromagnetic fields in a more general cylindrical spacetime; that is with both potentials $\psi$ and $\omega$. The second deals with the critical collapse of matter fields in cylindrical symmetry since there are few works on this subject in axisymmetric spacetimes. We intend to proceed with the collapse of massless scalar fields but relaxing the condition of hyperbolic canonicity of the line element.

\begin{acknowledgements}
 The authors thank the financial support of Brazilian agencies CNPq and CAPES.
\end{acknowledgements}

\section*{appendix}

First we define the auxiliary basis $\chi_k(y)$ as,

\begin{eqnarray}
\chi_k(y) = \frac{1}{2}\left(TL_{k+1}(y)+TL_k(y)\right)
\end{eqnarray}

\noindent and the basis functions $\Psi_k(y)$ and $\Phi_k(y)$ are,

\begin{eqnarray}
\Psi_k(y) &=& \frac{(2k^2+2k+3)}{(2k^2+6k+7)} \chi_{k+1}(y) + \chi_{k}(y) \nonumber \\
\\
\Phi_k(y) &=& \frac{(k+1)(2k+3)}{8 (k+2)(2k+5)} \chi_{k+2}(y) + \frac{(2k+3)}{8 (k+2)} \chi_{k+1}(y)+\nonumber \\
& & + \frac{(2k+3)}{8 (2k+1)} \chi_{k}(y)
\end{eqnarray}














\end{document}